\documentclass[twoside,12pt]{article}
\usepackage{times,color,verbatim,lscape,setspace,showlabels}
\usepackage{graphicx, amsmath, amssymb,color,threeparttable}
\usepackage[perpage,symbol*]{footmisc}

\usepackage[sort, compress]{natbib}

\def\boxit#1{\vbox{\hrule\hbox{\vrule\kern6pt
   \vbox{\kern6pt#1\kern6pt}\kern6pt\vrule}\hrule}}

\def\tit.arg{Semiparametric Small Area Estimation of Crop Acreage under Partially Linear Model}

\def\abst.arg{
We propose the semiparametric small area estimator under the partially linear model. Unlike the linear mixed models widely used in the small area estimation, which assume that the area effect is the random effect, we model the area effect as the unknown function of an area-indicative variable. We represent the nonparametric part for area effect using penalized splines, by which the estimation and inference are done in the linear mixed model framework. The mean-squared error of empirical estimators is shown and the testing for small area effect also considered. The agricultural application of this modeling shows our proposed method could get the reliable estimates. Additionally, some numerical simulations are demonstrated to express the good performance of this modeling.
}

\def\key.arg{
small area estimation;
partial linear model;
penalized spline regression.
}

\def\authorhere{
By
RONG ZHU$^1$, GUOHUA ZOU$^1$, HAIBIN XIE$^2$ and YI HU$^1$ \\

{\it $^1$Academy of Mathematics and Systems Science,
Chinese Academy of Sciences, Beijing {\rm 100190}, P. R. China}\\
{\rm  } \ \ {\rm rongzhu@amss.ac.cn}\\
{\it $^2$University of International Business and Economics,
Beijing {\rm 100029}, P. R. China}\\

}

\begin{document}

\begin{center}
{\large \bf \tit.arg} \vskip 3mm

\authorhere
\end{center}

\centerline{\small SUMMARY}
\abst.arg

\vskip 3mm\noindent
{\it Some key words:} \key.arg

\section{Introduction}
In the last decade, small area estimation has attracted increasing attention, as it's of interest to provide estimates for small subpopulation (small area) within the overall population of interest in many surveys. \cite{Rao:03} thoroughly reviewed the models and methods in his field, including the linear mixed models, generalized mixed models, and empirical \& hierarchical Bayesian methods.

In the linear models framework, the small area estimation models regularly assume the area effect, which is unknown, but as random effect to interpret the part of out-of linear mean, e.g., the area level linear model proposed by \cite{Fay:Herriot:79} and the unit level linear model by \cite{Battese:88}. For example, the traditional linear model for the unit-level is:
$$y_{ij}=\bold{x}_{ij}^T\bold{\theta}+v_i+e_{ij}, \ \ i=1,\cdots,m\eqno(1.1)$$
where $y_{ij}$ is the target variable for the $j$-th unit in $i$-th area, $\bold{x}_{ij}$ the auxiliary variables, $\bold{\theta}$ the fixed parameters, $v_i$ the random area-effect, $e_{ij}$ the model error.

However this random effect assumption is not suitable in application. We build the following unit-level model for the Agricultural Survey data at eastern Heilongjiang province in China, which is investigated by the National Bureau of Statistics of China in 2012.
$$y_{ij}=\beta_0+x_{ij}\beta_1+v_i+e_{ij}$$
where $y_{ij}$ and $x_{ij}$ are the bean acreage by survey and satellite remote sensing for the $j$th grid unit in $i$th District respectively, random effect $v_i\sim N(0,\sigma_v^2)$, and $e_{ij}$ the model error.

However when we analyze $$\tilde{v}_i=\bar{y}_i-\bar{x}_i\tilde{\beta}_1$$
where $\tilde{\beta}_1=[(x_{ij}-\bar{x}_i)^2]^{-1}[(x_{ij}-\bar{x}_i)(y_{ij}-\bar{y}_i)]$, it's shown in Fig.1 that
$\tilde{v}_i$ depends on $\bar{X}_i$, where $\bar{X}_i$ is average bean acreage in $i$th District. The plot is shown in Fig.1. And the correlation between them is 0.42, and if we make regression $\tilde{v}_i$ on $\bar{X}_i$ with intercept and know that the $\bar{X}_i$ effect on $\tilde{v}_i$ is highly statistically significant ($t$ value is 11.17, and $p<2\times 10^{-16}$). Thus it's not practical to assume $v_i$ is $v_i\sim N(0,\sigma_v^2)$  for this specific case. Furthermore it's an observation that it's not sufficiently reasonable to simply assume the linear relationship between $v_i$ and $\bar{X}_i$, as we use the smoothing spline to fit the function and get that there is the characteristics of flat head and tail in the curve.

\begin{figure}
\includegraphics[scale=0.9]{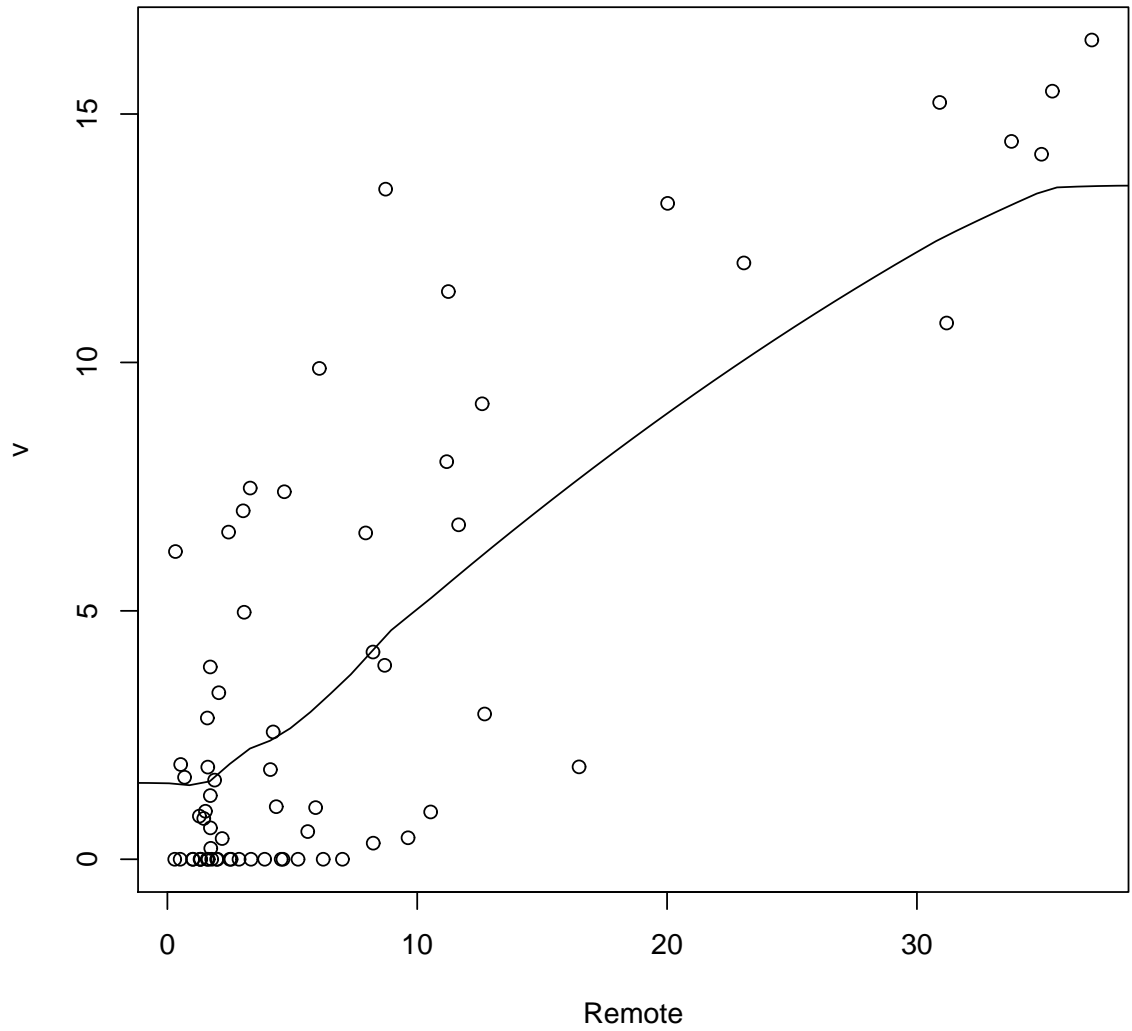}
\caption{The plot between $\tilde{v}_i$ and $\bar{X}_i$. Horizontal axis represents of $\bar{X}_i$, vertical axis $\tilde{v}_i$, and the solid line the curve fitted by smoothing spline.}
\end{figure}

Following the observation above, we proposed a semiparametric small area estimators under partially linear models.
We assume there exists an area-indicative variable, which can be used to explain the area effect, such as $\bar{X}_i$ in the agricultural survey case above. In fact, \cite{Opsomer:08} also studied the application using the hydrologic unit codes (HUCs) -specific effects.

Now, we assume $v_i$ is the function of the area-indicative variable $z_i$, that is $$v_i=f(z_i).\eqno(1.2)$$
Thus the traditional unit-level linear model is changed to a partial linear model, as follows:
$$y_{ij}=\bold{x}_{ij}^T\bold{\theta}+f(z_i)+e_{ij}, \ \ j=1,\cdots,n_i,\ \i=1,\cdots,m.\eqno(1.3)$$
For area-level linear model, the similar semiparametric model is also built as
$$y_i=\bold{x}_i^T\bold{\theta}+f(z_i)+e_i, \ \ i=1,\cdots,m\eqno(1.4)$$

Rather than assuming the random effect, we assume an unknown fuction of the area-indicative variable to represent the area effect in small area estimation. This proposed idea is different with the nonparametric small area estimation proposed by \cite{Opsomer:08} and further investigated by \cite{Salvati:10}, which built the nonparametric model for auxiliary variable.
While in this paper we abandon the random effect assumption and approximate the area effect by an unknown function.

Inspired by the application above, the aim of our paper is demonstrate the estimation and inference for model (1.3) and (1.4). However we just investigate model (1.3) among two models in this paper, as the methods for dealing with both models are identical and our application is the unit-level case. then we will go back this application in section 5. In section 2, the main method, penalized spline regression is briefly reviewed and then the estimation for our model by the $P$-spline is given. We investigate the mean-square error for the small area estimators and give the testing for the small area effect in section 3. Some finite sample performance is contained in section 4. Last we go back this agricultural case to investigate the application.

\section{Description of model and estimation}
Firstly, we describe the $P$-splines method, which has been the subject of detailed description in \cite{Ruppert:03} and \cite{Opsomer:08}.
Consider the simple model
$$t_i=m_0(z_i)+\varepsilon_i,\eqno(2.1)$$
where, $\{\varepsilon_i, i=1,\cdots,m\}$ are independent model errors with $\varepsilon_i\sim (0,\sigma^2)$, and $m_0(\cdot)$ is the unknown function, however, which can be estimated by $P$-splines.

Assume that the unknown function $m_0(\cdot)$ can be approximated sufficiently well by
$$m(z;\bold{\beta},\bold{\gamma})=\beta_0+z\beta_1+\cdots+z^p\beta_p+\sum\limits_{k=1}^K\gamma_k(z-\kappa_k)_{+}^p,\eqno(2.2)$$
where, $p$ is the degree of the spline, $(z)_{+}^p$ denotes truncates polynomial function $z^pI_{\{x>0\}}$, $\kappa_1<\cdots<\kappa_K$ the set of fixed knots, and $\bold{\beta}=(\beta_0,\cdots,\beta_p)T$ and $\bold{\gamma}=(\gamma_1,\cdots,\gamma_K)T$ are coefficient vectors.

The $P$-spline regression estimates are defined as the minimizers over $\bold{\beta}$ and $\bold{\gamma}$ of
$$\sum\limits_{i=1}^m\{t_i-m(z_i;\bold{\beta},\bold{\gamma})\}^2+\lambda\bold{\gamma}^T\bold{\gamma},\eqno(2.3)$$
where, $\lambda$ is a fixed penalty parameter.

Remark 1: The spline function (2.2) in fact uses the truncated polynomial spline basis $\{1,x,\cdots,x^p,(x-\kappa_1)_+^p,\cdots,(x-\kappa_K)_+^p\}$ to approximate the unknown function $m_0$;

Remark 2. Even for $p$ small (we set $p=1$ in our paper), the spline function can approximate sufficiently the smooth function $m_0(\cdot)$ with a high degree of accuracy, in which case we can ignore the lack-of-fit error $m_0(\cdot)-m(\cdot;\bold{\beta},\bold{\gamma})$, if provided the knots locations  are sufficiently spread out over the range of $z$ and $K$ is large enough.

Remark 3. The knots are often at equally spaced quantiles of the distributions of the covariate and $K$ is taken to be large relative to the size of the data set. And a typical knot choice for univariate $z$ may be one knot every four or five observations with a maximum number of 35-50.

Remark 4. Different $\lambda$ values result in the estimate of $P$-spline regression, so it's of interest to treat $\lambda$ as an unknown parameter, and treat the $\bold{\gamma}$ as a random-effect vector in the linear mixed model (LMM) specification, following which, $\bold{\beta}$, $\bold{\gamma}$ and $\lambda$ all can be estimated in the LMM framework. All the four remarks have been shown in detail in \cite{Ruppert:03}.

Finishing the $P$-spline description, now we turn back to our semi-parametric small area model:
$$y_{ij}=\bold{x}_{ij}^{T}\bold{\theta}+v(z_i)+\epsilon_{ij}$$
as the part of area effect, $v(z_i)$ can be estimated by using $P$-splines, the semi-parametric small area model can be approximated sufficiently well by splines regression.
Thus, it's reasonable to assume that the data follows the model
$$y_{ij}=\bold{x}_{ij}^{T}\bold{\theta}+z_i\beta+\bold{w}_i^{T}\bold{\gamma}+\epsilon_{ij},\eqno(2.4)$$
where, $\bold{w}_{i}=((z_i-\kappa_1)_{+}^p,\cdots,(z_i-\kappa_K)_+^p)^T$, and $\bold{\gamma}=(\gamma_1,\cdots,\gamma_K)^T$.
Or the model can be rewritten as
$$\bold{Y}=\bold{X}\bold{\theta}+\bold{Z}\beta+\bold{W}\bold{\gamma}+\bold\epsilon,\eqno(2.5)$$
where, $\bold{Y}=(\bold{Y}_1^T,\cdots,\bold{Y}_m^T)^T$ with $\bold{Y}_i=(y_{i1},\cdots,y_{in_i})^T$,
$\bold{X}=(\bold{x}_1^T,\cdots,\bold{x}_m^T)^T$ with $\bold{x}_i=(\bold{x}_{i1},\cdots,\bold{x}_{in_i})^T$,
$\bold{Z}=(z_1\bold{1}_{n_1}^T,\cdots,z_m\bold{1}_{n_m}^T)^T$, $\bold{W}=(\bold{w}_1\bold{1}_{n_1}^T,\cdots,\bold{w}_m\bold{1}_{n_m}^T)^T$, and
$\bold{\epsilon}=(\bold{\epsilon}_1^T,\cdots,\bold{\epsilon}_m^T)^T$ with $\bold{\epsilon}_i=(\epsilon_{i1},\cdots,\epsilon_{in_i})^T$.

The parameters $\bold{\gamma}$ can be treated as a random-effect vector following the the idea of $P$-splines, so we assume $$\bold{\gamma}\sim (\bold{0},\sigma_{\gamma}^2\bold{I}_K).\eqno(2.6)$$

Next, it's the task to estimate the parameters $\bold{\theta}$, $\beta$ and $\bold{\gamma}$ by BLUP (or EBLUP) theory under the linear mixed model specification.

If the variance components are known, the results can be obtained by the BLUP methods.
Let the fixed-effects parameters $\bold{\psi}=(\bold{\theta}^T,\beta)^T$, and design matrix $\bold{U}=(\bold{X} | \bold{Z})$, then the estimates of the fixed-effects parameters $\bold{\psi}$ and random-effect parameters $\bold{\gamma}$ are shown as follows:
$$\tilde{\bold{\psi}}=(\bold{U}^T\bold{V}^{-1}\bold{U})^{-1}\bold{U}^T\bold{V}^{-1}\bold{Y}\eqno(2.7)$$
$$\tilde{\bold{\gamma}}=\sigma_{\gamma}^2\bold{W}^T\bold{V}^{-1}(\bold{Y}-\bold{U}\tilde{\bold{\psi}}), \eqno(2.8)$$
where, $\bold{V}=\sigma_{\gamma}^2\bold{W}\bold{W}^T+\sigma^2\bold{I}$.
Furthermore, $$\tilde{\bold{\theta}}=(\bold{X}^T\bold{V}^{-1}\bold{X})^{-1}\bold{X}^T\bold{V}^{-1}\bold{Y}-(\bold{Z}^T\bold{M}\bold{Z})^{-1}(\bold{X}^T\bold{V}^{-1}\bold{X})^{-1}\bold{X}^T\bold{V}^{-1}\bold{Z}\bold{Z}^T\bold{M}\bold{Y}\eqno(2.9)$$
$$\tilde{\beta}=(\bold{Z}^T\bold{M}\bold{Z})^{-1}\bold{Z}^T\bold{M}\bold{Y}\eqno(2.10)$$
where,
$\bold{M}=\bold{V}^{-1}-\bold{V}^{-1}\bold{X}(\bold{X}^T\bold{V}^{-1}\bold{X})^{-1}\bold{X}^T\bold{V}^{-1}$.

In small area estimation, we are interested in predicting
$$\bar{Y}_i=\bar{\bold{X}}_i^T\bold{\theta}+z_i\beta+\bold{w}_i^T\bold{\gamma}\eqno(2.11)$$
for a given small area $i$, where, $\bar{\bold{X}}_i$ is the true population means of the auxiliary information. Therefore, we use $$\hat{\bar{Y}}_i=\bar{\bold{X}}_i^T\tilde{\bold{\theta}}+z_i\tilde{\beta}+\bold{w}_i^T\tilde{\bold{\gamma}}\eqno(2.12)$$
as a predictor of $\bar{Y}_i$.

If the variance components of unknown, the EBLUP estimates $\hat{\bold{\theta}}$, $\hat{\beta}$ and $\hat{\bold{\gamma}}$ are constructed by replacing $\sigma_{\gamma}^2$ and $\sigma^2$ by their estimators, which can be estimated by maximum likelihood (ML) or restricted maximum likelihood (REML) methods, and then get the small area EBLUP estimates
$$\bar{Y}_i^E=\bar{\bold{X}}_i^T\hat{\bold{\theta}}+z_i\hat{\beta}+\bold{w}_i^T\hat{\bold{\gamma}}\eqno(2.13)$$

\section{Theoretical properties}
\subsection{Prediction MSE}
Firstly the case of known variances is considered, that is, the prediction error $\hat{\bar{Y}}_i-\bar{Y}_i$.
We get
$$\hat{\bar{Y}}_i-\bar{Y}_i=\bold{b}_i(\tilde{\bold{\psi}}_i-\bold{\psi}_i)+\bold{w}_i\left(\sigma_{\gamma}^2\bold{W}^T\bold{V}^{-1}(\bold{Y}-\bold{U}\bold{\psi})-\bold{\gamma}\right),\eqno(3.1)$$
where, $\bold{b}_i=\bold{\bar{U}}_i-\bold{w}_i\sigma_{\gamma}^2\bold{W}^T\bold{V}^{-1}\bold{U}$.
Since the penalized spline method is estimated under the linear mixed effect models framework, the MSE of the small area estimators are calculated as:
$$E(\hat{\bar{Y}}_i-\bar{Y}_i)^2=\bold{b}_i(\bold{U}^T\bold{V}^{-1}\bold{U})^{-1}\bold{b}_i^T+\bold{w}_i\sigma_{\gamma}^2(\bold{I}-\sigma_{\gamma}^2\bold{W}^T\bold{V}^{-1}\bold{W})\bold{w}_i^T,\eqno(3.2)$$

In practice, often the variance components are estimated from the data under a few methods, and in this case we need to consider the MSE for EBLUP prediction error as follows:
$$\hat{\bar{Y}}_i^E-\bar{Y}_i=\hat{\bold{b}}_i(\hat{\bold{\psi}}_i-\bold{\psi}_i)+\bold{w}_i\left(\hat{\sigma}_{\gamma}^2\bold{W}^T\hat{\bold{V}}^{-1}(\bold{Y}-\bold{U}\bold{\psi})-\bold{\gamma}\right),\eqno(3.3)$$
where, $\hat{\bold{b}}_i=\bold{\bar{U}}_i-\bold{w}_i\sigma_{\gamma}^2\bold{W}^T\hat{\bold{V}}^{-1}\bold{U}$.

The approximations of MSE of EBLUP for the linear mixed models have been widely investigated: \cite{Kachar:Harville:84} studied a general approximation of MSE of EBLUP, to deal with underestimation problem, \cite{Prasad:Rao:90} derived the second-order approximation of MSE as the variances are estimated by moment methods under are level or unit level models, \cite{Datta:Lahiri:00} extended the Prasad-Rao approach to the case where ML or REML are used to estimate the variance components. Furthermore, \cite{Das:04} showed the rigorous derivation of the second-order approximation of MSE under the general linear mixed model, and \cite{Opsomer:08} applied the results of \cite{Das:04} to spline-based random component. In our paper, we also apply the the results of \cite{Das:04} and \cite{Opsomer:08}.

\textbf{Theorem 1} Suppose that:
(1) The following are bounded: $k$, $n_i$, the elements of $\bold{X}$, $\bold{Z}$ and $\bold{W}$, eigenvalues of $\bold{\Sigma}$, and the number of knots $K$;
(2) The true variance components $\sigma_{\gamma}^2$ and $\sigma^2$ are positive;

Then, the MSE of the EBLUP estimators is given as follows:
$$E(\hat{\bar{Y}}_i^E-\bar{Y}_i)^2=E(\hat{\bar{Y}}_i-\bar{Y}_i)^2+tr(\bold{S}\bold{V}\bold{S}^T\bold{\mathcal{I}}^{-1})+o(m^{-2}),\eqno(3.4)$$
where, $\bold{S}$ is a 2-size vector, with $S_1=\bold{w}_i\bold{W}^T\bold{V}^{-1}\left(\bold{I}_n-\sigma_{\gamma}^2\bold{W}\bold{W}^T\bold{V}^{-1}\right)$, $S_2=-\bold{w}_i\sigma_{\gamma}^2\bold{W}^T\bold{V}^{-1}\bold{V}^{-1}$, and $\bold{\mathcal{I}}$ is Fisher information matrix with respect to $(\sigma_{\gamma}^2,\sigma^2)$, whose elements $I_{ij}$ for $i,j=1,2$, equal $\frac{1}{2}tr(\bold{Q}\bold{B}_i\bold{Q}\bold{B}_j)$ for REML estimation,
or $tr(\bold{Q}\bold{B}_i\bold{Q}\bold{B}_j)-\frac{1}{2}tr(\bold{V}^{-1}\bold{B}_i\bold{V}^{-1}\bold{B}_j)$ for ML estimation, where, $\bold{B}_1=\bold{W}\bold{W}^T$, $\bold{B}_2=\bold{I}_n$, $\bold{Q}=\bold{V}^{-1}-\bold{V}^{-1}\bold{U}(\bold{U}^T\bold{V}^{-1}\bold{U})^{-1}\bold{U}^T\bold{V}^{-1}$.

Furthermore, the estimator of MSE is shown
$$mse(\hat{\bar{Y}}_i^E)=\hat{\bold{b}}_i(\bold{U}^T\hat{\bold{V}}^{-1}\bold{U})^{-1}\hat{\bold{b}}_i^T+\bold{w}_i\hat{\sigma}_{\gamma}^2(\bold{I}-\hat{\sigma}_{\gamma}^2\bold{W}^T\hat{\bold{V}}^{-1}\bold{W})\bold{w}_i^T+2(\bold{Y}-\bold{U}\hat{\bold{\psi}})^T\hat{\bold{S}}^T\hat{\bold{\mathcal{I}}}^{-1}\hat{\bold{S}}(\bold{Y}-\bold{U}\hat{\bold{\psi}}),\eqno(3.5)$$
where, the $\hat{\bold{S}}$, $\hat{\bold{\mathcal{I}}}$ are given by replacing the unknown variances in $\bold{S}$ and $\bold{\mathcal{I}}$ respectively. And the estimator of MSE is second-order approximated, that is, $$E[mse(\hat{\bar{Y}}_i^E)]=E(\hat{\bar{Y}}_i^E-\bar{Y}_i)^2+o(m^{-2})$$

The proof is shown in Appendix.

\subsection{Testing of Small Area Effect}
In our paper, the small area effect is assumed as an unknown function form of an area-indicative variable, so it's necessary to know if the small area effect reasonably exists. If without the area effect, then the use of simple synthesis estimators are both convenient and justified. The task then is to test the presence of the area effect in this subsection.

Both hypothesis are needed to test:

(H1)  the null hypothesis $H_{0,\beta}: \beta=0$ v.s. the alternative $H_{1,\beta}: \beta\in R$, and

(H2)  the null hypothesis $H_{0,\gamma}: \sigma_{\gamma}^2=0$ v.s. the one-sided alternative $H_{1,\gamma}: \sigma_{\gamma}^2>0$.

It's simple to test hypothesis H1 using $\chi^2$-test, as the form of the estimator $\tilde{\beta}$ is explicitly expressed.
Construct a test statistic $$T=\bold{Y}^T\bold{M}\bold{Z}(\bold{Z}^T\bold{M}\bold{Z})^{-1}\bold{Z}^T\bold{M}\bold{Y},\eqno(3.6)$$
which is $\chi_1^2$-distribution under the null hypothesis if assumed variances are known. When the variance components are unknown, the new test statistic $$\hat{T}=\bold{Y}^T\hat{\bold{M}}\bold{Z}(\bold{Z}^T\hat{\bold{M}}\bold{Z})^{-1}\bold{Z}^T\hat{\bold{M}}\bold{Y},\eqno(3.7)$$ inserting the estimated variances, is asymptotic $\chi_1^2$-distribution under the null hypothesis as the estimators are consistent.

The likelihood ratio test for one-sided testing shown in \cite{Self:Liang:87} is used to test hypothesis H2. Construct the likelihood ratio test statistic
$$LRT=\text{sup}_{\bold{\psi},\sigma_{\gamma}^2,\sigma^2}L(\bold{\psi},\sigma_{\gamma}^2,\sigma^2)-\text{sup}_{\bold{\psi},\sigma_{\gamma}^2=0,\sigma^2}L(\bold{\psi},\sigma_{\gamma}^2=0,\sigma^2),\eqno(3.8)$$
where, $L(\bold{\psi},\sigma_{\gamma}^2,\sigma^2)=-log|\bold{V}|-(\bold{Y}-\bold{U}\bold{\psi})^T\bold{V}^{-1}(\bold{Y}-\bold{U}\bold{\psi})$.
From \cite{Self:Liang:87} and \cite{Crainiceanu:Ruppert:04}, $LRT$ has the asymptotic distribution which is an equal mixture of a point mass at $0$ and a $\chi_1^2$-distributed, denoted $\frac{1}{2}\chi_0^2+\frac{1}{2}\chi_1^2$.

\section{Finite sample performance}

To evaluate the appropriateness of our proposed approach, we performed a limited simulation study under the our partial linear model (2.4). We took $\bold{\theta}=(1,1)^T$, $\bold{x}_{ij}=(1,x_{ij})^T$ with $x_{ij}$'s generated from the Uniform distribution on $[1/3,3]$,
$z_i$'s follows the Uniform distribution on $[1/2,2]$, each $n_i=4$ for all $i$, and $m=30,60, \text{or} \ \ 100$.

Five different models for the functions $v(z_i)$ were considered. The models were $\text{M}_1$: $v(z_i)=sin(z_i)$; $\text{M}_2$: $v(z_i)=1+z_i$; $\text{M}_3$: $v(z_i)=exp(z_i)$; $\text{M}_4$: $v(z_i)=\phi(z_i)$ where $\phi$ is the standard normal density function; $\text{M}_5$: $v(z_i)=1$.

We quantified the performances of estimators of mean-squared error by using empirical measures of relative bias and coefficient of variation. Relative bias of the mean-squared error estimator was defined to be the average of
$$\text{RB}_i=\frac{E(\text{mse}_i)-\text{SMSE}_i^0}{\text{SMSE}_i^0}, \eqno(4.1)$$
for $i=1,\cdots, m$, that is, $\text{RB}=\frac{1}{m}\sum\limits_{i=1}^m\text{RB}_i$, where $E(\text{mse}_i)$ was estimated empirically as the average of values of $\text{mse}_i$ over replicates (the replicate times $\text{B}=1,000$), and $\text{SMSE}_i^0$ defined as the average value of $\hat{\bar{Y}}_i^E-\bar{Y}_i^0$, where $\bar{Y}_i^0=\bar{\bold{X}}_i^T\bold{\theta}+v(z_i)$, over replicates that can be as the true value. And The coefficient of variation of the MSE estimator was taken to be
the average of
$$\text{CV}_i=\frac{[E(\text{mse}_i-\text{SMSE}_i)^2]^{1/2}}{\text{SMSE}_i^0}, \eqno(4.2)$$
for $i=1,\cdots, m$, that is, $\text{CV}=\frac{1}{m}\sum\limits_{i=1}^m\text{CV}_i$, where $E(\text{mse}_i-\text{SMSE}_i^0)^2$ was estimated empirically as the average of values of $(\text{mse}_i-\text{SMSE}_i^0)^2$ over replicates.

And we did the performances of testing the area effect for the five different models. The proportion (P) of rejecting the null hypothesis, no area effect, under $\alpha=0.05$ for 1,000 replicates.

Table 1 reports results in estimating the mean-squared error, and gives those of testing the area effect. Following it, we can observe that, for model M2 and M5, the estimators of mean-squared error by our method is very near with simulated mean-squared error with simulation error which is almost the same as the true value, and the difference is small for other models. This result indicates that the good small area estimators can be obtained by our method.
For testing the area effect, according to the five models, it's expected that the value of $P_2$ should be close to 0, and both $P_1$ and $P_2$ close to 0, which is identical with our simulation result. So the test statistic $\hat{\bold{T}}$ for testing the area effect is effective.

\section{Application}

Let us go back our agricultural survey case in the introduction section. In 2012, National Bureau of Statistics of China launched a project to estimate the crop acreage at Heilongjiang, a major agricultural province in China. Thanks to satellite remote sensing technology in agricultural survey, survey statisticians can get the values for each District, however the quality of data is poor and there may be relatively large gap between the real value and the remote sensing data because of the image recognition error. While survey data is accurate but it's often not practical to draw enough samples for each District to get the reliable estimates based on the survey data. Thus, the estimates can be improved by combining the satellite data and survey data.

Thanks to National Bureau of Statistics of China for providing the bean data, in this section we consider the prediction of bean acreage for 69 Districts in East Heilongjiang. The sample unit is the grid (1 grid=33.75 mu, where acreage measurement $1\text{mu}=150m\times150m=2.25\times10^4m^2$).  The sample size in each District is from 1 to 37, and the total sample size of all 69 Districts are 701.  Following the sample data analysis, we know the bean planting is a little rare and dispersed, as 435 (62\%) among 701 sample units have no bean acreage, while only 22 (32\%) District among 69 Districts have no bean. These description information about the samples are shown in Figure 2.

\begin{figure}
\includegraphics[scale=1.0]{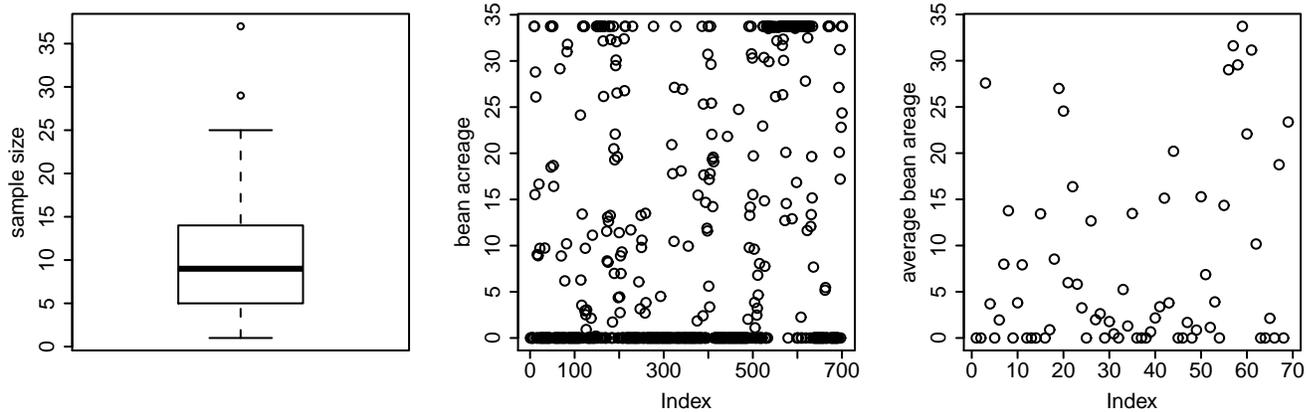}
\caption{The description of sample data. Left figure is the boxplot of sample size in each district, middle the plot of unit-level bean acreage, and right the plot of area(district)-level average acreage.}
\end{figure}

Because of the rareness and dispersion for bean acreage, the preliminary analysis shown in section 1 indicates it's not reasonable to assume the area effects $v_i\sim N(0,\sigma_v^2)$,
If the traditional nested linear model is used, the estimated parameters under the random effect linear model is:
$$\hat{y}_{ij}=4.46(0.86)+0.59(0.04)x_{ij}, \ \ \text{with}\ \ \hat{\sigma}_v^2=38.00, \hat{\sigma}_e^2=64.38$$
where ($\cdot$) means their std. error value and both p-values for intercept and $x$ are 0.

From this model results, it's noted that the variances are too large to get the reliable analysis.
Thus we proposed the partially linear model as shown in (1.3) and choose the average bean acreage based on the satellite data in each District ($\bar{X}_i$) as the area-indicative variable, which means that area effect $v_i$ can be seen as a unknown function of average bean acreage by the satellite data in each District, then get the small area estimates by $P$-spline.

Based on the results on remarks, we choose $p=1$ and knots are at equally quantiles of the distributions of $\bar{X}_i$ with $K=15$. And the Likelihood ratio test show it's statistically significant that $\sigma_{\gamma}^2>0$. Then the estimated parameters is gotten as follows:
$$\hat{y}_{ij}=1.61(0.46)+0.48(0.05)x_{ij}+0.25(0.05)\bar{X}_i, \ \ \text{with}\ \ \hat{\sigma}_{\gamma}^2=1.40, \hat{\sigma}_e^2=2.56$$
where ($\cdot$) means its std. error value and all p-values for three coefficients are less than $0.001$.

The small area estimation results are shown in Fig.3. Based on Figure 3, we know that the root mean squared error (RMSE) estimators of these estimates is so small relative to the estimates that we can sure these bean average acreage estimates are reliable based on the our assumed model. However there are two tricks for this result, one is that we ignore the model approximate error, though it's probably small relative to the estimation error, and the other is that the result is based on the model, which needs to be carefully considerable.

\begin{figure}
\includegraphics[scale=1.0]{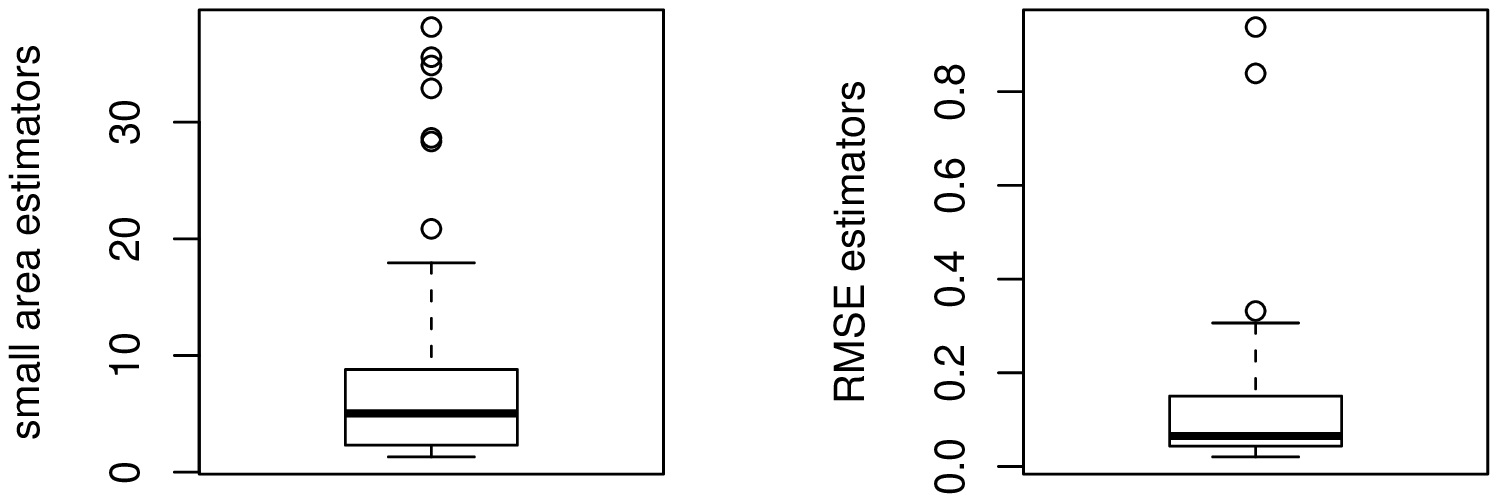}
\caption{The results of average bean acreage. Left figure is the average bean acreage estimators, and right root mean squared error estimators.}
\end{figure}

\section{Discussion}

Rather than assuming a random area effect for the part of out of linear mean in the small area estimation, we assume there exists an area-indicative variable in this paper, which can be used to explain the area effect, and the area effect is the unknown function of the area-indicative variable.
Then the estimators can be gotten by using penalized splines based on the partial linear model. Furthermore the mean-squared error of empirical estimators is shown, and testing for small area effect also considered.
Based on our application, we can know that this new proposition could get the reliable estimates.

However, in this paper there are some considerations which are worthy of further discussion and study.
We know that the good performance shown by the simulation is based on the fact that there is the area-indicative variable, which is not available sometimes. However, in many cases, we can find the area-indicative variable such as our application.

Following the works of \cite{Ruppert:03} and \cite{Opsomer:08}, penalized splines are made use to deal with our problem, but other semiparametric method could be applied here, such as the kernel method, which warrants our future research.

Additionally, to make the problem simple, we don't consider the the error caused by spline approximation, which is small relative to the estimation error according to\cite{Ruppert:03}.


\begin{table}
\renewcommand{\arraystretch}{1}
\begin{center}
\caption{Simulation results for 5 models with different $m$}
\begin{threeparttable}
\begin{tabular}{c | c | c c c c c}
\hline
 \text{Model} & $m$ & \text{SMSE} & \text{RB} & \text{CV} & $P_1$ & $P_2$\\
\hline\hline
M1 & 30 & 0.0425 & 0.176 & 0.758 & 0.395 & 1\\
 & 60 & 0.0228 & 0.227 & 0.347 & 0.218 & 0.998\\
 & 100 & 0.0147 & 0.248 & 0.342 & 0.348 & 0.995 \\
\cline{1-7}
M2 & 30 & 0.0179 & -0.0309 & 0.601 & 0.977 & 0.007\\
 & 60 & 0.0100 & -0.0548 & 0.786 & 0.998 & 0.008\\
 & 100 & 0.00588 & -0.0349 & 0.833 & 0.990 & 0.005\\
\cline{1-7}
M3 & 30 & 0.0389 & 0.466 & 0.505 & 0.218 & 0.956\\
 & 60 & 0.0202 & 0.379 & 0.567 & 0.182 & 0.0981\\
 & 100 & 0.0121 & 0.349 & 0.441 & 0.150 & 0.973\\
\hline
M4 & 30 & 0.0353 & -0.892 & 11.9 & 0.938 & 0.038\\
 & 60 & 0.0236 & -1.352 & 15.63 & 0.948 & 0.627\\
 & 100 & 0.017 & -0.139 & 7.07 & 0.852 &0.774\\
\hline
M5 & 30 & 0.0195 & -0.02034 & 0.9968 & 0.052 & 0.002\\
 & 60 & 0.00903 & 0.0176 & 0.236 & 0.042 & 0.004\\
 & 100 & 0.00561 & 0.0140 & 0.355 & 0.048 & 0.001\\
\hline
\end{tabular}
\begin{tablenotes}\footnotesize
\item{Note: $P_1$ means the frequency probability for testing the H1, and $P_2$ means testing the H2.}
\end{tablenotes}
\end{threeparttable}
\end{center}
\end{table}


\vskip 1cm

{\noindent}{\bf Acknowledgements}\\
RZhu's research was supported by National Natural Science Foundation of China (Grant nos.
11301514) and National Bureau of Statistics of China (Grant nos. 2012LZ012). GZou's research was supported by the National Natural Science Foundation of China (Grant nos.
11021161 and 70933003) and the Hundred Talents Program of the Chinese Academy of Sciences.

\baselineskip=18pt

\bibliographystyle{biometrika}
\bibliography{semi_ref}

\clearpage\pagebreak\newpage \thispagestyle{empty}

\section{Appendix}

\textbf{Proof of Theorem 1}

Because the model (2.5) is the special case for ANONA model, and \cite{Das:04} have given rigorous proofs for general mixed linear models and shown the details for the ANONA model, we just apply their result in the Appendix.

Denote $\bold{\delta}=(\sigma^2,\sigma_{\gamma}^2)^T$, $\delta_1=\sigma^2$, $\delta_2=\sigma_{\gamma}^2$, then $\hat{\bar{Y}}_i^E=\hat{\bar{Y}}_i(\hat{\bold{\delta}})$.

Under assuming normality of $\gamma$ and $e$, \cite{Kachar:Harville:84} showed that
$$\text{MSE}[\hat{\bar{Y}}_i(\hat{\bold{\delta}})]=\text{MSE}[\hat{\bar{Y}}_i(\bold{\delta})]+E[\hat{\bar{Y}}_i(\hat{\bold{\delta}})-\hat{\bar{Y}}_i(\bold{\delta})]^2,\eqno(A.1)$$
for any even and translation invariant estimator $\hat{\bold{\delta}}$, which is satisfied for ML or REML estimators, which are considered the ML and REML estimates for $\bold{\delta}$.

Let $\bold{D}_i=\sigma_{\gamma}^2\bold{W}^T\bold{V}^{-1}$, from Theorem 3.1 and Formula (3.4) in the paper of \cite{Das:04}, we get
$$E[\hat{\bar{Y}}_i(\hat{\bold{\delta}})-\hat{\bar{Y}}_i(\bold{\delta})]^2=tr\left[\left(\frac{\partial \bold{D}_i}{\partial\bold{\delta}}\right)^T\bold{V}\left(\frac{\partial \bold{D}_i}{\partial\bold{\delta}}\right)\mathcal{I}^{-1}\right]+o(d_*^{-2}),\eqno(A.2)$$
where $d_*=min\{d_1,d_2\}$ with $d_i$ for $i=1,2$ denoting the diagonal element of the information matrix associated with $\delta_i$ respectively.

For the REML method under the model (2.5), the restricted log-likelihood has the form
$$l_r(\bold{\delta})=c-1/2log|\bold{A}^T\bold{V}\bold{A}|-1/2\bold{Y}\bold{Q}\bold{Y},\eqno(A.3)$$
where, $c$ is a constant, $\bold{A}$ is any $n\times(n-k-1)$ matrix such that $\text{rank}(\bold{A})=n-k-1$ and $\bold{A}^T\bold{X}=\bold{0}$, $\bold{V}=\sigma_{\gamma}^2\bold{V}_1+\sigma^2\bold{V}_2$ with $\bold{V}_1=\bold{W}\bold{W}^T$, and $\bold{V}_2=\bold{I}$.
$\bold{Q}=\bold{A}(\bold{A}\bold{V}\bold{A})^{-1}\bold{A}^T=\bold{V}^{-1}-\bold{V}^{-1}\bold{U}(\bold{U}\bold{V}^{-1}\bold{U})^{-1}\bold{U}^T\bold{V}^{-1}$.
Then we get the REML estimates of $\delta_i$ is the solution of the equation
$$\frac{\partial l_r(\bold{\delta})}{\partial \bold{\delta}}=1/2\bold{Y}^T\bold{Q}\bold{V}_i\bold{Q}\bold{Y}-1/2tr(\bold{Q}\bold{V}_i)=0.\eqno(A.4)$$
$$\frac{\partial^2 l_r(\bold{\delta})}{\partial \delta_i\partial\delta_j}=1/2tr(\bold{Q}\bold{V}_i\bold{Q}\bold{V}_j)-\bold{Y}^T\bold{Q}\bold{V}_i\bold{Q}\bold{V}_j\bold{Q}\bold{Y},\eqno(A.5)$$
then $$\mathcal{I}_{reml}=-E\left(\frac{\partial^2 l_r(\bold{\delta})}{\partial \delta_i\partial\delta_j}\right)=1/2tr(\bold{Q}\bold{V}_i\bold{Q}\bold{V}_j).\eqno(A.6)$$

And for ML method, the log-likelihood has the form
$$l(\bold{\delta})=c-1/2log|\bold{V}|-1/2(\bold{Y}-\bold{U}\bold{\psi})^T\bold{V}^{-1}(\bold{Y}-\bold{U}\bold{\psi}),\eqno(A.7)$$
Then obtain the profile log-likelihood
$$l_p(\bold{\delta})=c-1/2log|\bold{V}|-1/2\bold{Y}\bold{Q}\bold{Y},\eqno(A.8)$$
by replacing the estimates of $\bold{\psi}$ in (A.5), following the ML estimates of $\bold{\delta}_i$ obtained from $\frac{\partial l_p(\bold{\delta})}{\partial \bold{\delta}}=0$., and Fisher Information matrix $$\mathcal{I}_{ml}=-E\left(\frac{\partial^2 l_p(\bold{\delta})}{\partial \delta_i\partial\delta_j}\right)=tr(\bold{V}\bold{V}_i\bold{V}\bold{V}_j)-1/2tr(\bold{Q}\bold{V}_i\bold{Q}\bold{V}_j).\eqno(A.9)$$

For ML and REML under the model (2.5), $d_*=min\{d_1,d_2\}$ with $d_1=[tr(Q^2)]^{1/2}=O(m)$ and $d_2=[tr(\bold{Q}\bold{W}\bold{W}^T\bold{Q}\bold{W}\bold{W}^T)]^{1/2}=O(m)$.
Thus we finished the proof of (3.4) in Theorem 1.

From the Theorem 4.1 in \cite{Das:04}, we get $$E[mse(\hat{\bar{Y}}_i^E)]=E(\hat{\bar{Y}}_i-\bar{Y}_i)^2+tr(\bold{S}\bold{V}\bold{S}^T\bold{\mathcal{I}}^{-1})+o(d_*^{-2})\eqno(A.10)$$
Thus, the proof of the second part of Theorem 1 is shown.


\end{document}